\documentclass[aps,prl,twocolumn,showpacs,superscriptaddress,amsmath,amssymb]{revtex4-1} 

\usepackage{amsmath,amssymb}
\usepackage{ulem}
\usepackage{color}
\usepackage[colorlinks,citecolor=blue]{hyperref}
\usepackage[capitalise]{cleveref}
\usepackage{graphicx}

\begin{document}
\title{A chiral Maxwell demon}
\author{Guillem Rossell\'o}
  \affiliation{Institut de F\'{i}sica Interdisciplinar i de Sistemes Complexos
  IFISC (CSIC-UIB), E-07122 Palma de Mallorca, Spain}
  \affiliation{ Instituto de Ciencia de Materiales de Madrid, CSIC, Cantoblanco, 28049 Madrid, Spain}
\author{Rosa L\'opez}
\affiliation{Institut de F\'{i}sica Interdisciplinar i de Sistemes Complexos
  IFISC (CSIC-UIB), E-07122 Palma de Mallorca, Spain}
\author{Gloria Platero}
\affiliation{ Instituto de Ciencia de Materiales de Madrid, CSIC, Cantoblanco, 28049 Madrid, Spain}
\date{\today}

\begin{abstract}
We investigate the role of chirality on the performance of a Maxwell demon implemented in a quantum Hall bar with a localized impurity. Within a stochastic thermodynamics description we investigate the ability of such a demon to drive a current against a bias. We show that the ability of the demon to perform is directly related to its ability to extract information from the system.  The key features of the proposed Maxwell demon are the topological properties of the quantum Hall system. The asymmetry of the electronic interactions felt at the localized state when the magnetic field is reversed joined to the fact that we consider energy dependent (and asymmetric) tunneling barriers that connect such state with the Hall edge modes allow the demon to properly work. 
\end{abstract}

\maketitle

\textit{Introduction.---}\label{*Introduction}  Since J. Clerk Maxwell envisioned in a \textit{gedanken} experiment the possibility of an entity, an intelligent agent (a demon to Lord Kelvin) capable of separating warm and cold particles of a gas without performing work apparently violating the second law of thermodynamics, the idea attracted plenty of theoretical attention \cite{Maruyama2009}. The apparent paradox was addressed on a basis in which information and entropy must be related \cite{Szilard1929}. Information is then, a physical magnitude and it fulfills physical laws \cite{Bennet1982}. Erasing information implies energy dissipation \cite{Landauer1961,Berut2012,Jun2014} that compensates the entropy reduction suffered by a system in the presence of the demon and ensures the validity of the second thermodynamic law. Such information-to-energy conversion is regarded as the solution of the Maxwell demon paradox \cite{Bennet1982}. Nowadays Maxwell's demon is regarded as a feedback control mechanism to convert information into energy.
Even though Maxwell's idea was enunciated as a \textit{gedanken} experiment, present technologies have made possible to build it at small scales, for instance by using Brownian particles \cite{Toyabe2010, Roldan2014}, single electrons \cite{Koski2014} and lasers pulses \cite{Vidrighin2016}. However, demons for quantum systems \cite{Lloyd1997} are hard to experimentally design and work and thus show a scarcer experimental activity due to technical difficulty of implementing a truly quantum demon\cite{Pekola2015,Toyabe2010} despite the possibility of improved performance \cite{Horowitz2014}.  It is worth mentioning that the great progress in the development of stochastic thermodynamics, has resulted in theoretical proposals for stochastic Maxwell demons 
\cite{Esposito2010,Schaller2011,Esposito2012,Strasberg2013,Strasberg2014,Kutvonen2015,Pekola2016} and finally in an experimental implementation of an autonomous Maxwell demon using coupled quantum dots \cite{Koski2015}.  \\
A key ingredient for the performance of stochastic Maxwell demons is the breakdown of detailed balance conditions \cite{Esposito2012} as direct consequence of a feedback mechanism. The breakdown of such relations is usually achieved through an asymmetry in the system, e.g. in the tunneling barriers of quantum dot systems \cite{Esposito2012,Strasberg2013}. Breaking the local detailed balance (LDB) condition creates an imbalance between forward and backward processes from which the demon can profit.

An extra degree of freedom for the demon to work is found in topological systems.  Here, asymmetries, either kinetic (from tunneling events) or electrostatic  appear naturally in quantum Hall platforms due to the chirality of the edge states \cite{Sanchez2005,Sanchez2009}. The out-of-equilibrium LDB is broken in quantum Hall (QH) systems with localized impurities due to a non-symmetric electrostatic response of the system when the  magnetic field is reversed \cite{Lopez2012}. Precisely we benefit from this feature to devise a Maxwell demon feedback scheme in a QH device with an impurity.  We show how the chirality of the edge channels favors the operation of a demon that pushes an electrical current against a bias. In the demon protocol the drag of the electrical current opposite to the bias direction occurs by means of information-to-energy conversion as we demonstrate. The protocol requires two conditions to be satisfied \textit{(i)} electrostatic interactions must be asymmetric when the magnetic field or edge motion is reversed and \textit{(ii)} tunneling events between the edge modes and the localized level must be energy dependent. Under these circumstances the demon is able to convert information into work to drag an electrical current that moves contrary to the applied bias. Below we describe the theoretical framework for the chiral demon protocol.

\begin{figure}[!ht]
\includegraphics[width=\linewidth]{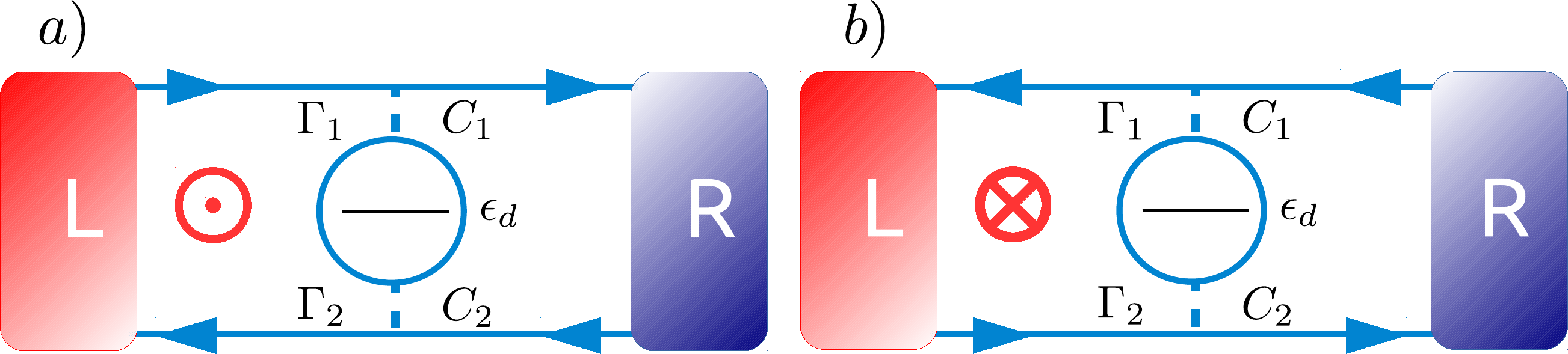}
\caption{\small The device composed by two electronic reservoirs (Left and Right), the QH bar with edge states represented with lines with arrows (representing direction of transport) and the localized impurity of energy level $\epsilon_d$ with tunnel couplings $\Gamma_{1(2)}$ and capacitive couplings $C_{1(2)}$. In a) (for $B>0$) the upper (lower) edge state travels from L(R) to R(L), instead in b) (for $B<0$) the upper (lower) edge state travels from R(L) to L(R).}
\label{QHMD}
\end{figure}

\textit{Theoretical approach---}\label{*Theoretical model} As mentioned our device is a topological setup  consisting of a single quasi-localized state of energy $\epsilon_d$ and it can only be singly occupied since we restrict ourselves to the Coulomb blockade regime with a sufficient strong Coulomb energy. Such level is tunnel coupled to chiral edge states of a QH system at filling factor $\nu=1$, see \cref{QHMD}. The applied magnetic field defines the direction of transport (represented by arrows) in each of the chiral modes. Tunneling events between chiral conducting states, $i=1,2$, and the quasi-localized state are modeled by the tunneling rates $\Gamma_{i}^{0(1)}$ that depend on energy. Thus, tunneling rates are different depending on the dot charge state, either empty (0) or filled (1). Electrons in the quasi-localized state are capacitively coupled to those in the edge channels via the capacitances $C_{i}$ \cite{Sanchez2004,Sanchez2005,Sanchez2009}. Due to the chirality of the edges states, the coupling of the quasi-bound state to the left and right reservoirs changes by reversing the magnetic field that exchanges the direction of motion of such modes, see \cref{QHMD}.  In this manner, for $B>0$ [\cref{QHMD}(a)] the level is coupled to the left (L) reservoir through the upper capacitance ($C_1$) and to the right (R) reservoir through the bottom capacitance ($C_2$). Under these considerations the quasi-localized level has a chemical potential given by
$
\mu_d^+= \epsilon_d + \frac{e}{C_1+ C_2} \left(C_1 V_L + C_2 V_R \right),
$
being $V_{L(R)}$ the voltage in reservoir $L(R)$ Changing the field direction  ($B<0$), [\cref{QHMD}(b)] leads to a reverse motion for the  conducting modes, with a chemical potential 
$
\mu_d^-= \epsilon_d + \frac{e}{C_1+ C_2} \left(C_2 V_L + C_1 V_R \right).
$
Note that the difference between both chemical potentials for $B<0$, and $B>0$ becomes
\begin{equation}\label{deltamu}
 \mu_d^-- \mu_d^+=e \eta (V_R-V_L),
\end{equation} 
where $\eta=\frac{C_1- C_2}{C_1+ C_2} $.  Such difference is null unless the setup is under non-equilibrium conditions and the capacitances are not symmetric, i.e. $\eta \neq 0$. Electrostatic interactions  are not symmetric when $B$ is reversed \cite{Sanchez2004}.

To describe transport through the interacting level we employ the Master 
equation framework taking advantage of the fact that in the Coulomb Blockade regime transport happens sequentially. We permit two charge states, namely $|0\rangle$ and $|1\rangle$ whose occupation probabilities are governed by the master equations
\begin{align}\label{Mastereq}
\dot{p}_0 = -(W_{10}^L+W_{10}^R) p_0 + (W_{01}^L+W_{01}^R) p_1,  \\
\dot{p}_1 = -(W_{01}^L+W_{01}^R) p_1 + (W_{10}^L+W_{10}^R) p_0. \nonumber
\end{align}
The transition rates between those two states are given by $W_{mn}^\alpha$ where $m$ is the final state of the dot, $n$ is the initial state and, $\alpha$ is the reservoir to/from which the electron comes. Due to the demon protocol, described below and which needs to be taken into account in the master equation, the transition rates that fill the dot ($W_{10}^{\alpha}$) are always given for a situation in which $B>0$ and those that empty the dot ($W_{01}^{\alpha}$) occur for $B<0$. This is the basis for our Maxwell demon feedback scheme to work properly since both transitions rates are not related to each other by a LDB condition\cite{Lopez2012} as we show below. 
These previous transition rates, in the sequential tunneling regime,  are readily obtained from Fermi's golden rule and read:
\begin{align}\label{Trates}
W_{10}^{L(R)} &= \Gamma_{1(2)}^0 f\left(\mu_d^{+}-\mu_{L(R)}\right),\\
W_{01}^{L(R)} &= \Gamma_{1(2)}^1 \left[1- f\left(\mu_d^{-}-\mu_{L(R)}\right) \right],
\end{align}
being $f(\mu_d^{+/-}-\mu_{L(R)})=1/(1+\exp\beta[\mu_d^{+/-}-\mu_{L(R)}])$ the Fermi distribution function of the reservoir $L(R)$ with electrochemical potential $\mu_{L(R)}$ and $\beta=1/k_B T$. As we remarked before, these rates depend on the orientation of the magnetic field through their dependence on the quasi-bound state chemical potential $\mu_d^{+/-}$.
Due to the chirality of the system,  the inversion of the magnetic field changes the quasi-bound dot energy at which transport is more favored and breaks the LDB as 
$
\frac{W_{10}^{L}}{W_{01}^{L}} \sim  \frac{\Gamma_1^0}{\Gamma_1^1}   e^{-\beta \left( \epsilon_d - \frac{e\Delta V}{2} \right)} \left(1- \left[1 - 2 f \left(\epsilon_d - \frac{e\Delta V}{2} \right) \right] \beta \eta \frac{e\Delta V}{2}\right),
$
being $\Delta V=V_L-V_R$ the applied bias. 
We also observe that an energy dependent model for the tunneling rates $\Gamma_1^0 \neq \Gamma_1^1$ breaks the LDB even if capacitances are equal. However, we show here that the demon works properly and drags electricity against bias voltage, solely when both requirements  are met \textit{(i)} asymmetric capacitances, and \textit{(ii)} energy  dependent tunneling rates with asymmetric barriers.
We take the tunneling rate energy dependence to be within the WKB approximation. Here,  the energy dependence of the tunneling rates is exponential, as shown experimentally \cite{MacLean2007}
\begin{align}
\Gamma_\alpha^{0(1)} = \Gamma_\alpha e^{k_\alpha (\mu_d^{+/-}-E_\alpha)},
\end{align}
where $\alpha=1,2$ denotes the barrier, $k_\alpha$ models the energy dependence and $E_\alpha$ is the top energy of the barrier. We profit from the fact that in this approximation barriers are asymmetric whenever $k_1 \neq k_2$.

\begin{figure}[!ht]
\includegraphics[width= \linewidth]{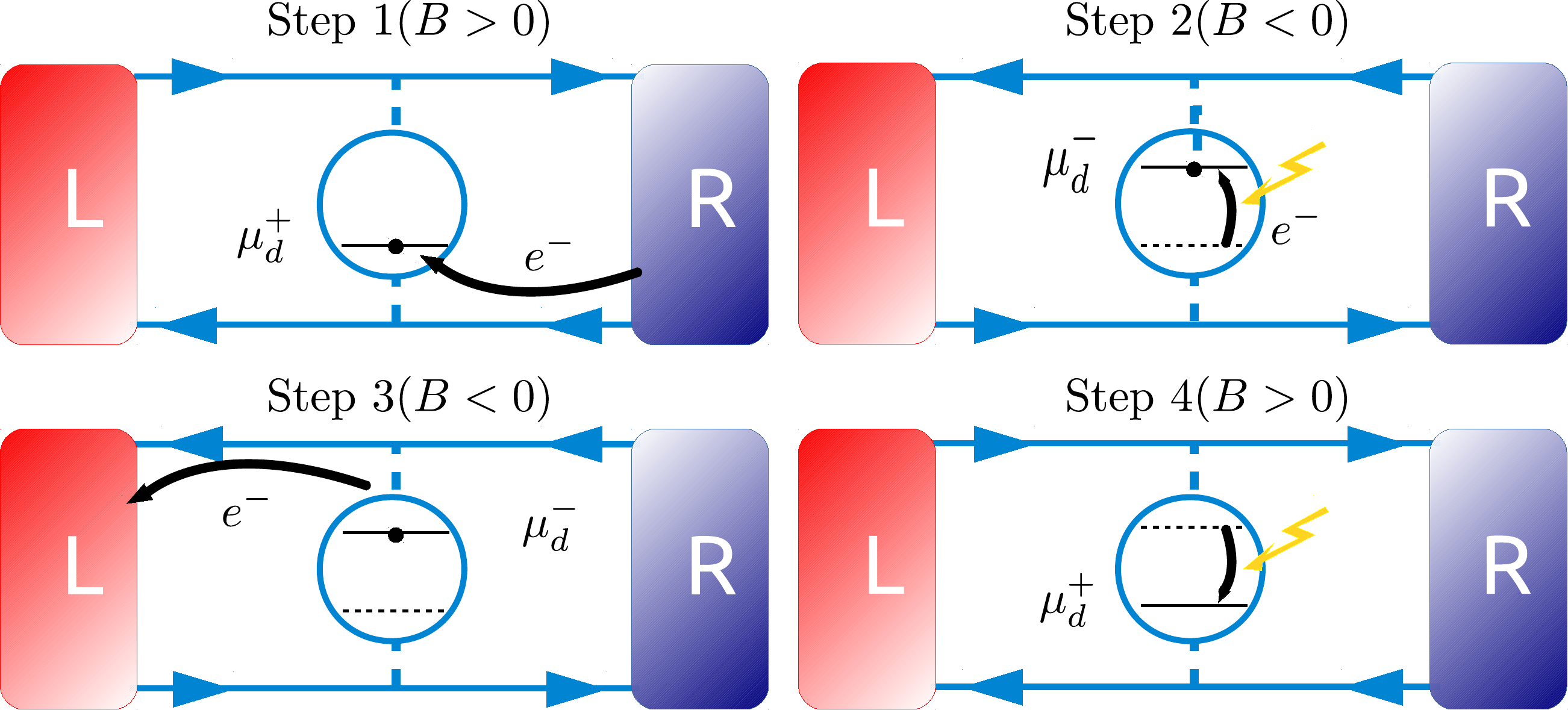}
\caption{\small Sketch of the performed action by the demon to push electrons against a bias. The demon's action is represented by yellow lightning (steps 2 and 4). }
\label{fig:demon}
\end{figure}

\textit{Demon protocol.---}\label{*Working Principle} As explained before, LDB is broken, either because of the energy dependence of the barriers or due to the distinct orientation of the magnetic field. Under these circumstances  we devise a working process for the demon (see \cref{fig:demon}) and investigate if it would be able to drive a current against a voltage bias ($V_L>V_R$), i.e. to drive a current from right to left. The process is as follows:


\textbf{Step 1:} Starting with $B>0$ and an empty dot, the process is triggered when an electron enters the dot from the right contact (probability $\propto \Gamma_2^0$). Due to the orientation of the magnetic field, the energy of the electron in the dot is 
$\mu_d^+$. 

\textbf{Step 2:} The demon detects that the dot is singly occupied and accordingly it changes the direction of the magnetic field to $B<0$, this raises the level energy for the localized state up to $ \mu_d^-$.

\textbf{Step 3:} Since now the energy of the electron inside the localized state has augmented it becomes easier to tunnel out through the upper barrier (probability $\propto \Gamma_1^1$) and as a consequence the dot is emptied. This leads to transport of one electron from the right to the left reservoir even though $V_L>V_R$.

\textbf{Step 4:} To finish, the demon detects  the change in the  localized state occupation and restores the initial magnetic field to $B>0$.
 
Note that the demon pushes electrons against a bias, this operation is optimal as long as the localized state level energy is increased as much as possible therefore helping charges climb against the applied bias. The energy difference after reversing the magnetic field is encountered in \cref{deltamu}. Then, we need to take $\eta<0$, i.e. $C_2>C_1$, given that $V_L>V_R$. Therefore the feedback scheme is more effective whenever $\eta$ is close to $-1$ or at the very nonlinear regime (large $V_L-V_R$).

Although tunneling processes in Steps 1 and 3 are against the bias and thus less likely, this inconvenient can be overcome by cleverly engineering the tunneling rates, see below and \cref{kratio}.

To implement theoretically the demon in our transport description, all these steps need to be incorporated in the master equation description \cref{Mastereq}. Then, the  feedback scheme is taken into account by taking the transition rates at the correct magnetic field orientations.

\textit{Results.---}\label{*Results} In order to characterize transport and to see whether the demon is able to drive a current against the applied bias or not, we calculate the currents using the probabilities from \cref{Mastereq}: $I_L=e(-W_{10}^L p_0 + W_{01}^L p_1 )$. By this definition currents are positive when particles go into a reservoir and negative otherwise and fulfill the particle conservation relation $I_L+I_R=0$. Then, solving \cref{Mastereq} in the steady state regime, the probabilities are easily obtained, yielding

\begin{equation}\label{Icurrent}
I_L=\frac{e(W_{01}^L W_{10}^R-W_{10}^L W_{01}^R)}{W_{01}^R+W_{01}^L+W_{10}^R+W_{10}^L}.
\end{equation}

We see from this expression that the first term accounts for particles entering the dot from the right and exiting through the left (positive contribution) and the second term accounts for particles entering the dot from the left and leaving through the right (negative contribution). In order to improve performance of the Maxwell demon we want the first term to be bigger than the second one since we want a current flowing from right to left. 

From the expressions of the transition rates and the current, \cref{Trates,Icurrent} respectively, we see that to favor the first term of the current we need to enhance $\Gamma_{1}^1 \Gamma_{2}^0$ with respect to $\Gamma_{1}^0 \Gamma_{2}^1$. From the expressions of the tunneling rates we obtain their ratio:

\begin{equation}\label{kratio}
\frac{\Gamma_{1}^1 \Gamma_{2}^0}{\Gamma_{1}^0 \Gamma_{2}^1}=e^{(k_1-k_2)(\mu_d^--\mu_d^+)}.
\end{equation}

It is observed then that we are able to favor the numerator by taking $k_1>k_2$, given that $\mu_d^--\mu_d^+ \geq 0$. 

\begin{figure}[!ht]
\centering
\includegraphics[width= 0.95\linewidth]{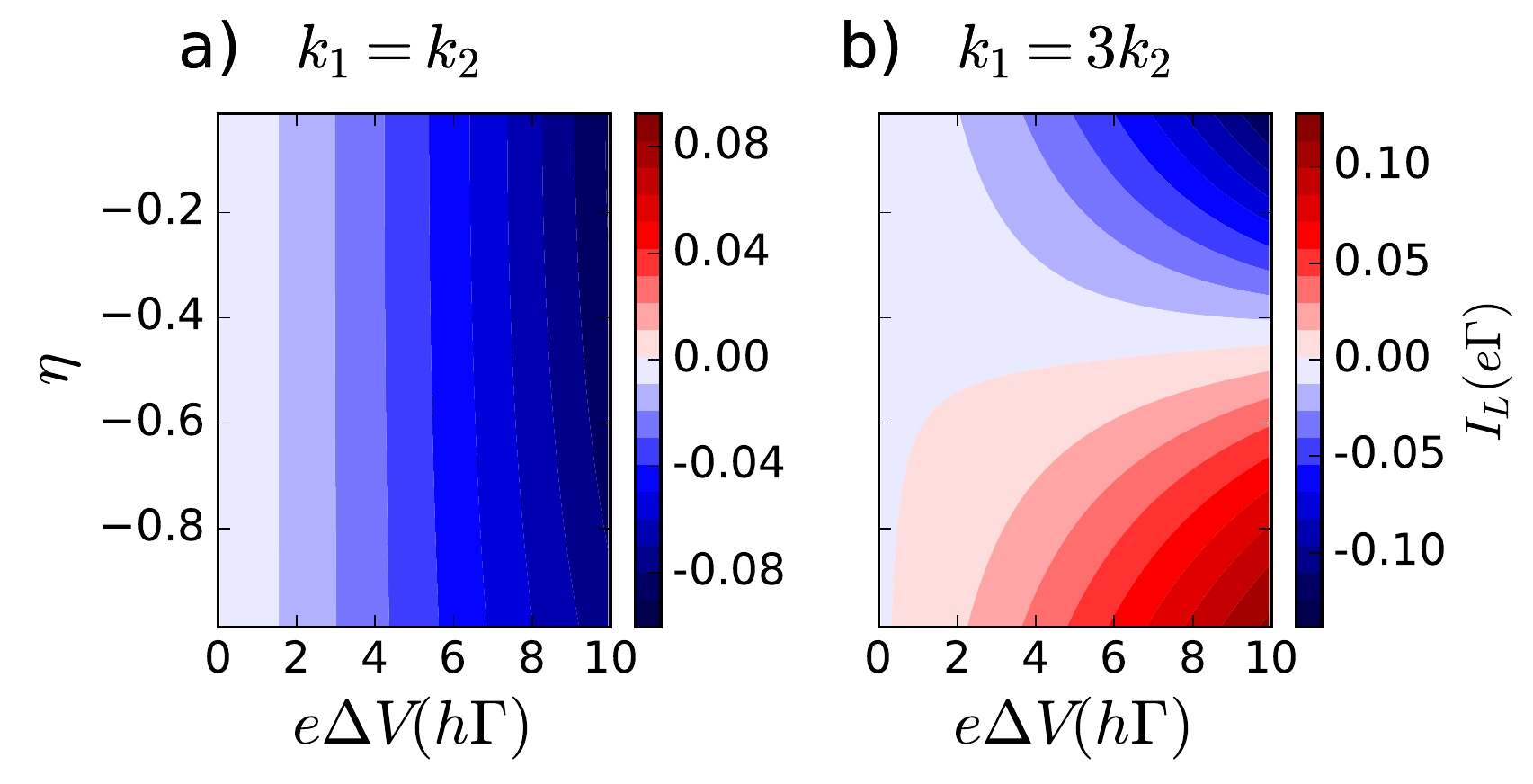}
\caption{\small Current to (positive) or from (negative)  the left reservoir $I_L$ for different values of $\eta$ as a function of $\Delta V=V_L-V_R$. The current is negative when it flows in the bias voltage direction. The parameters have been chosen so that $k_1=k_2$ in (a) and $k_1=3 k_2$ in (b). We have taken $\Gamma_1=\Gamma_2=\Gamma$, $k_1= 0.1 /h \Gamma$ and, $k_B T = 10 h\Gamma$.}
\label{IL}
\end{figure}

In \cref{IL} we represent the charge current in reservoir $L$. Current is negative as long as it flows with the bias according to our sign criterion. In \cref{IL} (a) we observe that symmetric barriers lead to a current that always follows the bias independently of feedback strength $\eta$. However, in \cref{IL} (b) when the tunneling barriers are different and for a sufficiently high $\eta$ value the current starts to flow opposite to the bias, becoming positive. This notable fact is thanks to the action of the demon. Importantly,  the action of the demon, i.e. its ability to change the energy level inside the dot $\mu^--\mu^+\propto \eta \Delta V$, then by increasing the applied voltage the current flowing against the bias enhances as well. The bias which in principle is an obstacle that needs to be overcome becomes rather a help because the action of the demon augments with it.

\begin{figure}[!ht]
\centering
\includegraphics[width= 0.95\linewidth]{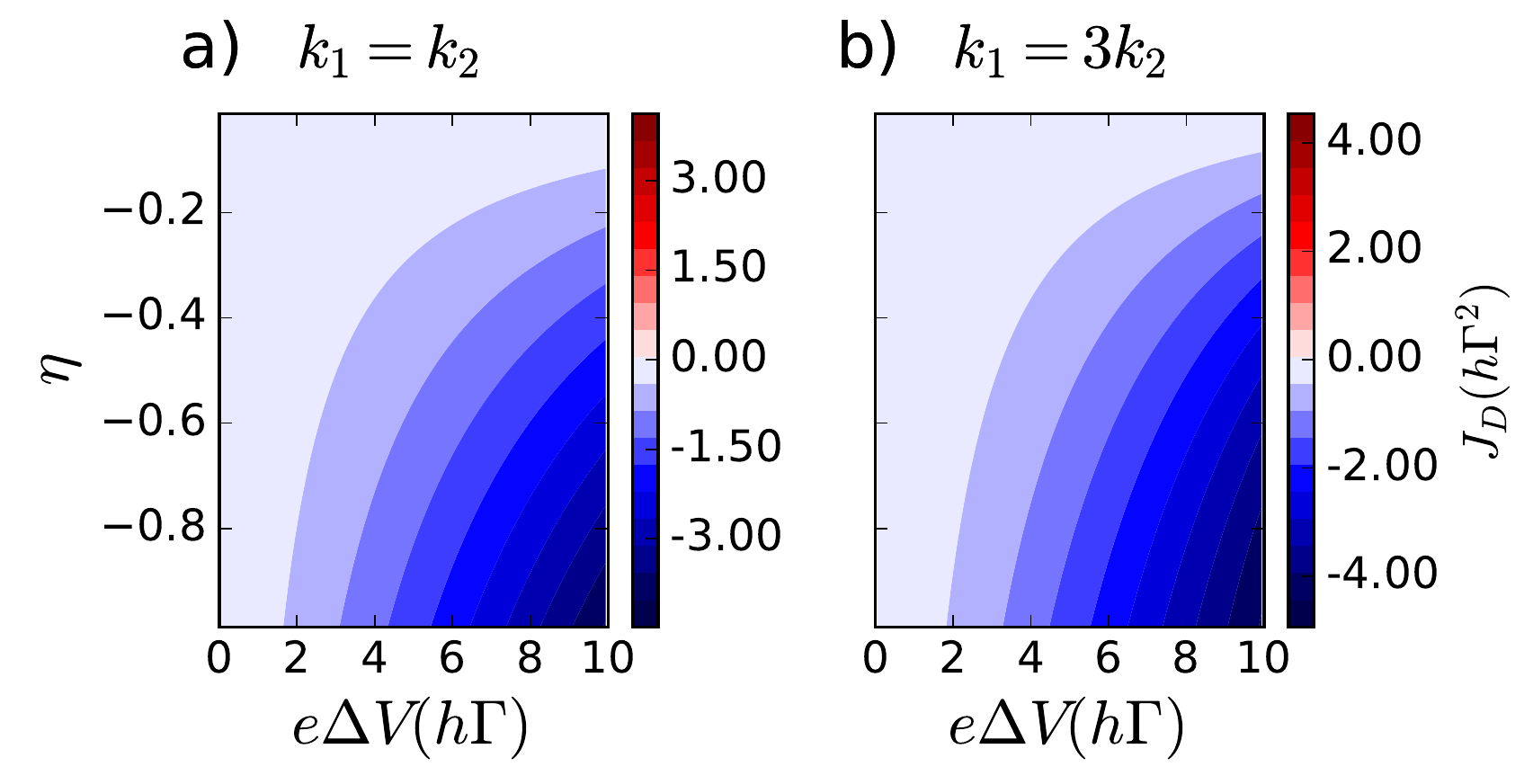}
\caption{\small Energy current ($J_D$) injected by the demon for different values of $\eta$. The parameters have been chosen so that $k_1=k_2$ in (a) and $k_1=3 k_2$ in (b). Parameters are the same as in \cref{IL}.}
\label{JD}
\end{figure}

To fully characterize the demon and since it is not ideal, the energy flow from the system to the demon and the demon entropy information flow are evaluated. The energy current is related to the energy change that the demon causes on the electrochemical potential of the localized impurity. We calculate this current by measuring the number of electrons going in and out of the dot times their respective energies, yielding 

\begin{align}\label{eqJD}
J_D=(\mu_d^+-\mu_d^-)\frac{(W_{10}^{L}+W_{10}^{R})(W_{01}^{L}+W_{01}^{R})}{W_{10}^{L}+W_{10}^{R}+W_{01}^{L}+W_{01}^{R}}.
\end{align}
Hence, we see that this energy current is proportional to the  energy difference of the electrochemical potentials caused by the demon times the activity current which measures how many particles go in and out of the impurity. Our findings  for the energy current carried by the demon are shown in \cref{JD}.  The energy injected by the demon increases with $\eta$  and with the applied voltage, as expected. Comparing the results for symmetric barriers [\cref{JD} (a)] and asymmetric ones  [\cref{JD} (b)] we show that the  current energy does not change significantly. This indicates that the energy injected by the demon although present and necessary is not the key factor in pushing the current against the bias. Let us now move on to the information entropy.

The final step in the characterization of the demon is the key quantity of an ideal Maxwell demon, i.e. the information that it can extract from the system in order to perform \cite{Esposito2012}. 
We start from the system information \cite{Esposito2010} given by Shannon's entropy 
$
S= -k_B \sum_m p_m \ln p_m,
$
where $p_m$ are the occupation probabilities given by \cref{Mastereq}. Then the entropy balance, $\dot{S}=\dot{S}_e+\dot{S}_i$, can be written as an entropy production $\dot{S}_i$ and an entropy flow $\dot{S}_e$ that satisfy $\dot{S}_i= -\dot{S}_e$.

The entropy flow in the system can be separated in the standard form of the entropy flow given by the exchanged heat in each reservoir ($\nu$) over the temperature and an extra term accounting for the entropy flow to the demon $\dot{S}_D$:

\begin{equation}\label{SeIF}
\dot{S}_e =  \sum_{\nu=L,R} \frac{J_{(\nu)}}{T}+\dot{S}_D.
\end{equation}
The term $\dot{S}_D$ consists of two parts, the entropy flow caused by the energy flux $J_D$ and an information flow $I_F$ that powers the demon.
This second term labeled $I_F$ is the information flow extracted by the demon in its application of the feedback protocol: 
\begin{equation}
I_F=\dot{S}_D-\frac{J_D}{T}.
\end{equation}
This information current is the main characteristic of the Maxwell demon since it measures the information that the demon extracts from the system in order to be able to apply the necessary feedback scheme. Specially in the case of an ideal Maxwell demon ($J_D=0$), it is the only measurable quantity associated to the Maxwell demon \cite{Esposito2012}.

\begin{figure}[!ht]
\centering
\includegraphics[width= 0.95\linewidth]{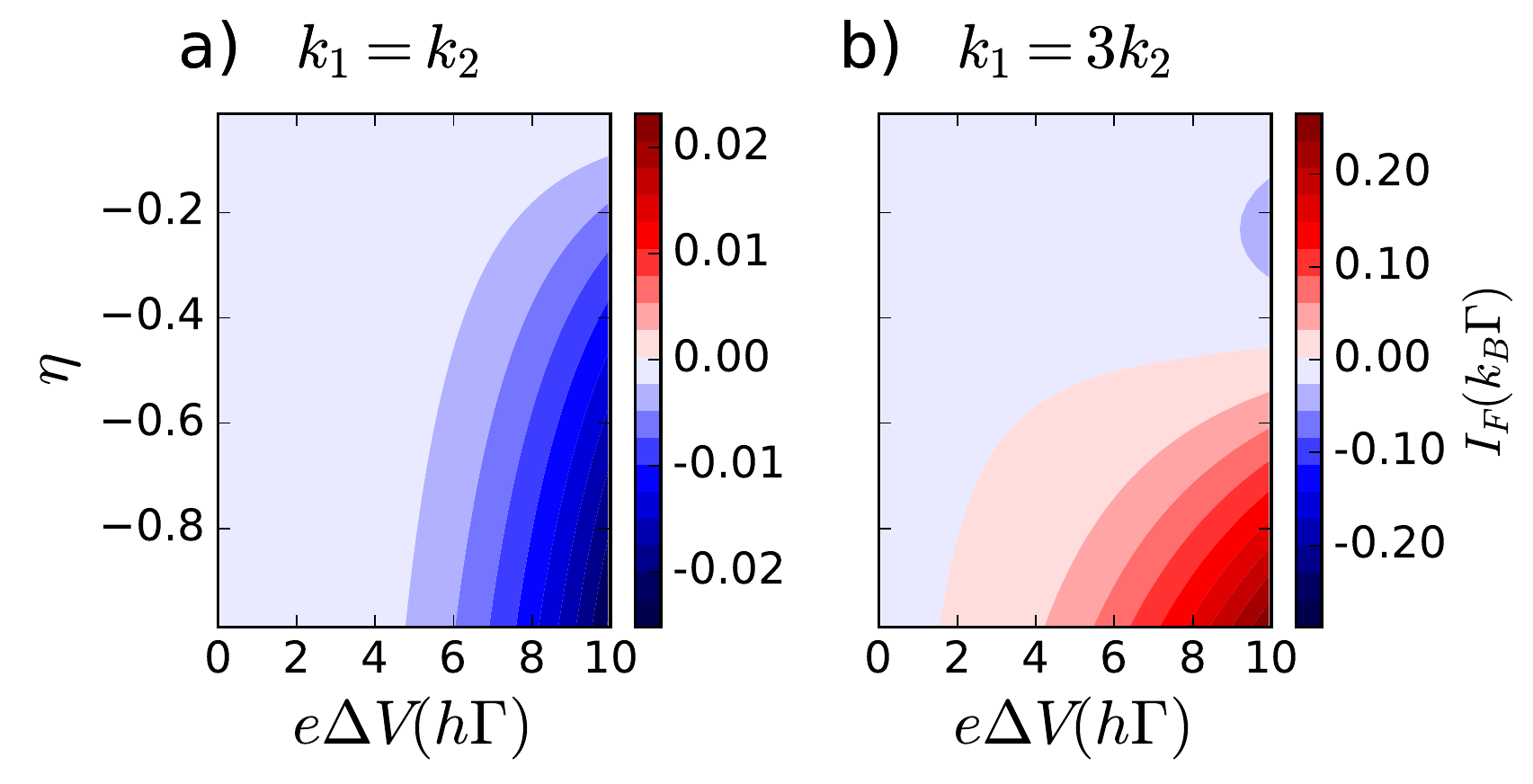}
\caption{\small  Information current ($I_F$) of the demon for different values of $\eta$. The parameters have been chosen so that $k_1=k_2$ in (a) and $k_1=3 k_2$ in (b). Parameters are the same as in \cref{IL}.}
\label{IF}
\end{figure}

 \cref{IF} represents the information current for the demon. In \cref{IF} (a) we observe that the information current is always negative meaning that information is flowing from the demon to the system. This corroborates that the demon is acting wrongly, being unable to extract information from the system. However, when barriers depend differently on energy [see \cref{IF} (b)] the demon  starts to be able to extract information from the system. Noticeably, we observe that the cases in which the demon drags an electrical current against the bias system coincide with the cases in which the demon is able to extract information from the system, compare \cref{IL,IF}.  This is the definitive indicator that the extraction of information is the key to our demon's operation. 

\textit{Conclusions.---}\label{*Conclusions} We have studied the effect of a non-ideal Maxwell demon feedback on transport through a localized state in a quantum Hall system. 
We proposed a working principle for a Maxwell demon based on the chirality of the edge states. We showed that it is applicable under the conditions: \textit{(i)} asymmetry of the capacitive interactions as the magnetic field is reversed, under non equilibrium conditions and \textit{(ii)} energy dependence of the tunneling probabilities through the barriers.
We show with a precise feedback protocol how a demon is able to push electrical current against the applied bias voltage by extracting information from the system. We demonstrate that this is effectively the case, the demon works satisfactorily whenever is able to extract  information from the system. Our scheme opens an important avenue for the design of quantum Maxwell demons that benefit from the topological properties of quantum matter in interacting systems.


\begin{acknowledgments}
G.R and R.L. were supported by MINECO Grants No. FIS2014-2564. G.P. and G.R. were supported by MINECO through Grant No. MAT2014-58241-
P. G.R. acknowledges financial support from the ``Red nacional de sistemas fuera del equilibrio".
\end{acknowledgments}

\bibliographystyle{apsrev4-1}
\bibliography{MaxwellDemon}

\end{document}